\documentclass[twoside,a4paper]{articlek}

\renewcommand{\refname}{REFERENCES}
\def\be{\begin{equation}}
\def\ee{\end{equation}}

\def\BibTeX{{\rm B\kern-.05em{\sc i\kern-.025em b}\kern-.08em
            T\kern-.1667em\lower.7ex\hbox{E}\kern-.125emX}}

\usepackage{graphicx}
\begin{document}
\sloppy
\twocolumn[{

{\large\bf FLUCTUATIONS OF ELECTROMAGNETIC FIELD AT THE INTERFACE BETWEEN
MEDIA}\\

{\small Ladislav \v{S}amaj, Ladislav.Samaj@savba.sk,
Institute of Physics, SAS,
D\'ubravsk\'a cesta 9, 845 11 Bratislava, Slovakia \\
Bernard Jancovici, Bernard.Jancovici@th.u-psud.fr,
LPT, Universit\'e de Paris-Sud, Orsay, France}\\
}]

\section{1. INTRODUCTION}
This study is related to the fluctuation theory of electromagnetic (EM)
fields, charges, and currents in systems formulated in the three-dimensional 
(3D) Cartesian space of points ${\bf r}=(x,y,z)$.
The studied problem is inhomogeneous along one of the axis, say along 
the first coordinate $x$, and translationally invariant along the plane formed
by the remaining two coordinates normal to $x$, denoted as ${\bf R}=(y,z)$.
The model consists of two semi-infinite media (conductor, dielectric or vacuum)
with the frequency-dependent dielectric functions $\epsilon_1(\omega)$ and
$\epsilon_2(\omega)$ which are localized in the complementary half-spaces
$x>0$ and $x<0$, respectively (see Fig. 1).
The interface between media is the plane $x=0$.
We assume that the media have no magnetic structure and the magnetic
permeabilities $\mu_1=\mu_2=1$.
The media and the radiated EM field are in thermal equilibrium
at some temperature $T$, or the inverse temperature $\beta=1/(k_{\rm B}T)$
with $k_B$ being the Boltzmann constant.

\begin{figure} [h,t]
\begin{center} 
\includegraphics[width=80mm,clip]{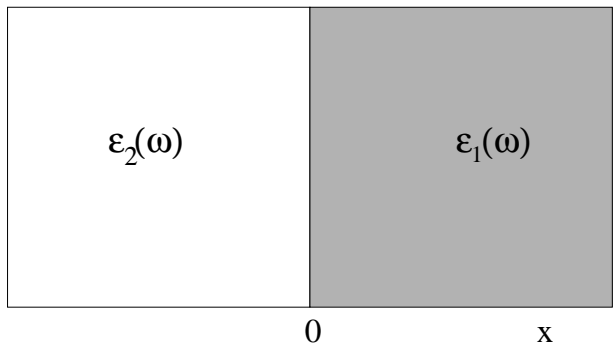}
\end{center}
\vspace{-2mm}
\caption{Two semi-infinite media characterized by the dielectric functions
$\epsilon_1(\omega)$ and $\epsilon_2(\omega)$.}
\end{figure}

The different electric properties of the media give rise to a surface 
charge density $\sigma(t,{\bf R})$ at time $t$ and at a point 
${\bf r}=(0,{\bf R})$ on the interface which must be understood as being 
the microscopic volume charge density integrated along the $x$-axis on 
some microscopic depth from the interface.
In thermal equilibrium, this quantity fluctuates in time around
its mean value, usually equal to 0.
Based on the elementary electrodynamics, the surface charge density is 
associated with the discontinuity of the normal $x$ component of 
the electric field at the interface: In Gauss units, we have
\be \label{1}
4\pi\sigma(t,{\bf R }) = E_x^+(t,{\bf R}) - E_x^-(t,{\bf R}) ,
\ee
where the superscript $+$ ($-$) means approaching the interface through
the limit $x\to 0^+$ ($x\to 0^-$).
The tangential $y$ and $z$ components of the electric field are
continuous at the interface.

The fluctuations of the surface charge density at two points on the interface, 
with times different by $t$ and distances different by $R=\vert {\bf R}\vert$, 
are correlated due to the EM interactions among the charged 
particles forming the two media. 
We are interested in the (symmetrized) two-point surface charge correlation
function defined by
\be \label{2}
S(t,{\bf R}) \equiv \frac{1}{2} \langle \sigma(t,{\bf R }) \sigma(0,0)  
+ \sigma(0,0) \sigma(t,{\bf R }) \rangle^{\rm T} , 
\ee
where $\langle \cdots \rangle^{\rm T}$ represents a truncated equilibrium 
statistical average, $\langle AB\rangle^{\rm T} = \langle A B \rangle
- \langle A \rangle \langle B \rangle$, at the inverse temperature $\beta$.
With regard to the relation (\ref{1}), the function (\ref{2}) is related
to fluctuations of the EM field on the interface between media.
Although the system is not in the critical state, due to the presence of 
long-ranged EM forces in the media the asymptotic large-distance 
behavior of the surface charge correlation function (\ref{2}) 
exhibits in general a long-ranged tail of type
\be \label{3}
\beta S(t,{\bf R}) \sim \frac{h(t)}{R^3} , \qquad R\to\infty ,
\ee 
where the form of the prefactor function $h(t)$ depends on
``the physical model'' used.
We can consider: 
\begin{itemize}
\item
Either {\em classical} or {\em quantum} mechanics.
According to the correspondence principle, the two theories
provide the same results in the high-temperature limit $\beta\hbar\to 0$.
\item
Either {\em non-retarded} or {\em retarded} regime of the EM interaction.
In the non-retarded (non-relativistic) regime, the speed of light $c$ is 
taken to be infinitely large, $c=\infty$, ignoring in this way magnetic 
forces acting on the particles.
In the retarded (relativistic) regime, $c$ is assumed finite and so
the particles are fully coupled to both electric and magnetic parts of 
the radiated EM field.
\end{itemize} 
The particles themselves are non-relativistic, classical or quantum.

It is useful to introduce the Fourier transform
\be \label{4}
S(t,{\bf q}) = \int {\rm d}^2 R\, \exp(-{\rm i}{\bf q}\cdot {\bf R})
S(t,{\bf R}) ,
\ee
with ${\bf q}=(q_y,q_z)$ being a 2D wave vector.
Since, in the sense of distributions, the 2D Fourier transform
of $1/R^3$ is $-2\pi q$, a result equivalent to (\ref{3}) is that
$\beta S(t,{\bf q})$ has a kink singularity at ${\bf q}={\bf 0}$,
behaving like
\be \label{5}
\beta S(t,q) \sim -2\pi h(t) q, \qquad q\to 0 .
\ee

The correlation function $S(t,{\bf R})$ and its long-ranged decay 
(\ref{3}) are of importance for two reasons.
Firstly, the Fourier transform is a measurable quantity by scattering
experiments.
The fact that $S(t,q)$ is linear in $q$ for small $q$ makes this quantity 
very different from the usual correlation functions with short-ranged decay 
which are proportional to $q^2$ in the limit $q\to 0$.
The second application arises in the context of the average polarization 
${\bf P}$ of a charged system, localized inside a domain $\Lambda$ of volume 
$\vert\Lambda\vert$ entoured say by vacuum and exposed to a constant electric 
field ${\bf E}$. 
In the linear limit, we have
\be \label{6}
\frac{P_j}{\vert\Lambda\vert} = \sum_k \chi_{jk} E_k 
\qquad j,k=x,y,z,
\ee
where $\chi_{jk}$ is the dielectric susceptibility tensor.
One expects that an extensive quantity like ${\bf P}$ divided by the
volume $\vert\Lambda\vert$ tends, for a fixed temperature, to a constant 
in the thermodynamic limit $\vert\Lambda\vert\to\infty$, 
independently of the shape of the infinite domain $\Lambda$.
This is not true in the case of the dielectric susceptibility tensor.
It can be shown that $\chi_{jk}$ contains contributions from both short-ranged 
bulk and long-ranged surface charge (\ref{3}) correlation functions which 
causes its shape dependence \cite{Choquard89,Jancovici04}.
Such a phenomenon is predicted by the macroscopic laws of electrostatics
\cite{LL1}.

The two media configuration studied so far was restricted to a conductor,
localized say in the half-space $x>0$ with the dielectric function
$\epsilon_1(\omega)\equiv \epsilon(\omega)$, in contact with vacuum
of the dielectric constant $\epsilon_2(\omega)=1$.
The obtained results can be summarized as follows.

$\bullet$ {\bf Classical non-retarded regime:}
Let the conductor be modelled by a classical Coulomb fluid composed
of charged particles with only the instantaneous Coulomb interactions.
By a microscopic analysis \cite{Jancovici82}, the long-ranged decay 
of the static (i.e. $t=0$) surface charge correlation was found such
that
\be \label{7}
h_{\rm cl}^{(\rm nr)}(0) = - \frac{1}{8\pi^2} ,
\ee
where the subscript ``cl'' means ``classical'' and the upperscript
``nr'' means ``non-retarded''.
The same result has been obtained later \cite{Jancovici95} by simple
macroscopic arguments based on a combination of the linear response
theory and the electrostatic method of images.
Note the universal form of $h_{\rm cl}^{(\rm nr)}(0)$, independent of
the composition of the classical Coulomb fluid.
 
$\bullet$ {\bf Quantum non-retarded regime:}
The quantum description of a general conductor is very complicated. 
A simplification arises in the so-called jellium model, i.e. 
a system of identical pointlike particles of charge $e$, mass $m$
and bulk number density $n$, immersed in a uniform neutralizing
background of charge density $-e n$.
The dynamical properties of the jellium have a special feature:
There is no viscous damping of the long-wavelength plasma oscillations
for identically charged particles \cite{PN}.
The frequencies of non-retarded nondispersive long-wavelength collective
modes, namely $\omega_p$ of the bulk plasmons and $\omega_s$ of the
surface plasmons, are given by
\be \label{8}
\omega_p = \left( \frac{4\pi ne^2}{m} \right)^{1/2}, \qquad
\omega_s = \frac{\omega_p}{\sqrt{2}} .
\ee
The dielectric function of the jellium is well described by a simple
one-resonance Drude formula
\be \label{9}
\epsilon(\omega) = 1 - \frac{\omega_p^2}{\omega(\omega+{\rm i}\eta)} ,
\ee
where the dissipation constant $\eta$ is taken as positive infinitesimal,
$\eta\to 0^+$.
The prefactor function $h(t)$ for the quantum jellium was obtained in 
Refs. \cite{Jancovici85a,Jancovici85b}. 
It has the nonuniversal form
\be \label{10}
h_{\rm qu}^{(\rm nr)}(t) = - \frac{1}{8\pi^2} 
\left[ 2 g(\omega_s) \cos(\omega_s t) - g(\omega_p) \cos(\omega_p t) \right] ,
\ee
where
\be \label{11}
g(\omega) = \frac{\beta\hbar\omega}{2} 
\coth\left( \frac{\beta\hbar\omega}{2} \right)
\ee
and the subscript ``qu'' means ``quantum''.
In the high-temperature limit $\beta\hbar\to 0$, the function
$g(\omega)=1$ for any $\omega$ and the quantum formula (\ref{10})
reduces to the classical non-retarded one
\be \label{12}
h_{\rm cl}^{(\rm nr)}(t) = - \frac{1}{8\pi^2} 
\left[ 2 \cos(\omega_s t) - \cos(\omega_p t) \right] .
\ee
For $t=0$, we recover the classical static result (\ref{7}).

The present study reviews some new results in the field 
\cite{SJ08,JS09a,JS09b}.
These results concern a generalization of the formalism to contacts between 
all kinds of media, i.e. conductors, dielectrics and vacuum, and to 
the retarded regime.
It turns out that, for any time $t$, the inclusion of retardation effects 
surprisingly converts the quantum results into the static classical ones.  

To deal with the physical problem of such complexity, we use a macroscopic
theory of equilibrium thermal fluctuations of EM field, initiated by
Rytov \cite{Rytov53} and further developed in Ref. \cite{LR}; we adopt 
the notation from the course by Landau and Lifshitz \cite{LP} (see Section 2).
Based on Rytov's fluctuational theory, we derive in Section 3 the general 
formula for the surface charge correlation function between two
semi-infinite media.
The analysis of the asymptotic large-distance form of this formula,
in both non-retarded and retarded regimes, is the subject of Section 4. 
We first consider the special configuration of the jellium conductor 
in contact with vacuum (Subsection 4.1) and then the general configuration
(Subsection 4.2).
Section 5 is a Conclusion. 

\section{2. FLUCTUATIONAL ELECTRODYNAMICS}
Let us consider a non-magnetoactive $(\mu=1)$ medium with the isotropic,
frequency and (possibly) position dependent, dielectric function
$\epsilon(\omega;{\bf r})$.
The coupled EM field is defined, in the classical format, by the scalar
potential $\phi(t,{\bf r})$ and the vector potential ${\bf A}(t,{\bf r})$.
In the considered Weyl gauge $\phi=0$, the microscopic electric and
magnetic fields are given by
\be \label{13}
{\bf E} = - \frac{1}{c} \frac{\partial {\bf A}}{\partial t} , \qquad
{\bf B} = {\rm curl}\, {\bf A} .
\ee
The elementary excitations of the {\em quantized} EM field are
described by the photon operators $\hat{A}_j$ $(j=x,y,z)$ which are
self-conjugate Bose operators; $\hat{A}_j(t,{\bf r})$ will denote
the vector-potential operator in the Heisenberg picture. 
The construction of all types of photon Green's functions is based
on the retarded Green function $D$, defined as the tensor
\be \label{14}
i D_{jk}(t;{\bf r},{\bf r}') = 
\langle \hat{A}_j(t,{\bf r}) \hat{A}_k(0,{\bf r}') -
\hat{A}_k(0,{\bf r}') \hat{A}_j(t,{\bf r}) \rangle 
\ee
for $t>0$ and equal to $0$ for $t<0$.
The Fourier transform in time of the retarded Green function reads
\be \label{15}
D_{jk}(\omega;{\bf r},{\bf r}') = \int_0^{\infty} d t\,
e^{i\omega t} D_{jk}(t;{\bf r},{\bf r}') .
\ee
For media with no magnetic structure, the Green function tensor possesses
the important symmetry
\be \label{16}
D_{jk}(\omega;{\bf r},{\bf r}') = 
D_{kj}(\omega;{\bf r}',{\bf r}) 
\ee

The EM fields, being in thermal equilibrium with medium, 
are random variables which fluctuate around their mean values
obeying macroscopic Maxwell's equations.
To deal with thermal fluctuations, Rytov put a classical current 
${\bf j}(t,{\bf r})$ due to the thermal motion of particles into 
the medium \cite{Rytov53}. 
This current acts as an ``external force'' on the vector-potential 
operator in the interaction Hamiltonian
\be \label{17}
{\cal H}_{\rm int}(t) = - \frac{1}{c} \int d^3 r\,
{\bf j}(t,{\bf r}) \cdot \hat{\bf A} .
\ee
The mean values of the components of the vector-potential operator
can be expressed in terms of the retarded Green function
by using Kubo's linear response in currents:
\be \label{18}
\frac{\bar{A}_j(\omega,{\bf r})}{c} = - \frac{1}{\hbar c^2} 
\int d^3 r'\, \sum_k D_{jk}(\omega;{\bf r},{\bf r}') 
j_k(\omega,{\bf r}') .
\ee 
The fact that the mean value $\bar{\bf A}$ satisfies the macroscopic
Maxwell equation due to the classical current ${\bf j}$ implies 
the differential equation of the dyadic type fulfilled by 
the retarded Green function tensor:
\begin{eqnarray}
\sum_{l=1}^3 \left[ \frac{\partial^2}{\partial x_j \partial x_l}
- \delta_{jl} \Delta - \delta_{jl} \frac{\omega^2}{c^2} 
\epsilon(\omega;{\bf r}) \right] D_{lk}(\omega;{\bf r},{\bf r}') 
\nonumber \\
= - 4 \pi \hbar \delta_{jk} \delta({\bf r}-{\bf r}') . \label{19}
\end{eqnarray} 
Here, in order to simplify the notation, the vector ${\bf r}=(x,y,z)$ 
is represented as $(x_1,x_2,x_3)$.
The differential equation must be supplemented by certain boundary 
conditions.
The second space variable ${\bf r}'$ and the second index $k$ only
act as parameters, the boundary conditions are formulated with respect 
to the coordinate ${\bf r}$ and the Green function 
$D_{lk}(\omega;{\bf r},{\bf r}')$ is considered as a vector
with the components $l=x,y,z$.
There is an obvious boundary condition of regularity at infinity,
$\vert {\bf r}\vert \to \infty$.
At an interface between two different media, the boundary conditions
correspond to the macroscopic requirements that the tangential
components of the fields ${\bf E}$ and ${\bf H}={\bf B}$ be continuous.
The vector potential is related to the Green function tensor
via the linear response (\ref{18}), and the electric and magnetic fields 
are related to the vector potential by (\ref{13}). 
This is why the role of the vector components $E_l$, up to an irrelevant
multiplicative constant, is played by the quantity
\be \label{20}
i \frac{\omega}{c} D_{lk}(\omega;{\bf r},{\bf r}')
\ee
and the role of the vector component $H_l$ is played by the quantity
\be \label{21}
\sum_j {\rm curl}_{lj} D_{jk}(\omega;{\bf r},{\bf r}') .
\ee
Here, we use the notation
${\rm curl}_{lj} = \sum_m e_{lmj}\partial/\partial x_m$ with
$e_{lmj}$ being the unit antisymmetric pseudo-tensor. 
For our geometry in Fig. 1, the tangential components (\ref{20}) and 
(\ref{21}), which are continuous at the interface $x=0$, correspond to 
indices $l=y,z$.

Eq. (\ref{18}) defines $-D_{jk}(\omega;{\bf r},{\bf r}')/(\hbar c^2)$ 
as the tensor of generalized susceptibilities corresponding to 
the variable ${\bf A}/c$.
The fluctuation-dissipation theorem tells us that the fluctuations
of random variables can be expressed in terms of the corresponding
susceptibilities.
For the assumed symmetry (\ref{16}), the theorem implies that
\be \label{22}
[ A_j({\bf r}) A_k({\bf r}') ]_{\omega} = - \coth( \beta\hbar\omega/2 )\,  
{\rm Im}\, D_{jk}(\omega;{\bf r},{\bf r}') ,
\end{equation} 
where the spectral distribution $[ A_j({\bf r}) A_k({\bf r}') ]_{\omega}$ 
is the Fourier transform in time of the symmetrized (truncated)
correlation function
\begin{equation} \label{23}
\frac{1}{2} \left\langle \hat{A}_j(t,{\bf r}) \hat{A}_k(0,{\bf r}')
+ \hat{A}_k(0,{\bf r}') \hat{A}_j(t,{\bf r}) \right\rangle^{\rm T} .
\end{equation}
The spectral distribution of the electric-field fluctuations can be 
easily found by using the first relation in Eq. (\ref{13}),
\begin{eqnarray} 
[ E_j({\bf r}) E_k({\bf r}') ]_{\omega} & = &   
\frac{\omega^2}{c^2} [ A_j({\bf r}) A_k({\bf r}') ]_{\omega} 
\nonumber \\
& = & - \frac{\omega^2}{c^2} \coth(\beta\hbar\omega/2) \nonumber \\ 
& & \qquad \times {\rm Im}\, D_{jk}(\omega;{\bf r},{\bf r}') . \label{24}
\end{eqnarray}

\section{3. SURFACE CHARGE CORRELATIONS}
The retarded Green function tensor for the studied problem of 
two semi-infinite media in Fig. 1 is obtained as the solution of
the dyadic differential Eq. (\ref{19}) supplemented by the mentioned
boundary conditions \cite{SJ08}.
Since the system is translationally invariant in the ${\bf R}$-plane 
perpendicular to the $x$ axis, we introduce the Fourier transform of 
the retarded Green function tensor with the wave vector ${\bf q}=(q_y,q_z)$,
\be \label{25}
D_{jk}(\omega;{\bf r},{\bf r}') = \int \frac{{\rm d}^2q}{(2\pi)^2}
{\rm e}^{{\rm i}{\bf q}({\bf R}-{\bf R}')} D_{jk}(\omega,{\bf q};x,x') .
\ee
Let us define for each of the half-space regions the inverse length
$\kappa_j(\omega,q)$ ($j=1,2)$ by
\be \label{26}
\kappa_j^2(\omega,q) = q^2 - \frac{\omega^2}{c^2} \epsilon_j(\omega) ,
\qquad {\rm Re}\, \kappa_j(\omega,q) > 0 ;
\ee
from two possible solutions for $\kappa_j$ we choose the one with
the positive real part in order to ensure the regularity of
the Green function at asymptotically large distances from the interface.
We shall only need the quantity $D_{xx}(\omega,q;x,x')$ for which
the obtained results can be summarized as follows:

(i) If $x,x'>0$,
\begin{eqnarray} 
D_{xx} & = & \frac{4\pi\hbar c^2}{\omega^2\epsilon_1} \delta(x-x')
- \frac{2\pi\hbar (cq)^2}{\omega^2\epsilon_1\kappa_1} \Bigg(
{\rm e}^{-\kappa_1\vert x-x'\vert} \nonumber \\ & &
\quad + \frac{\kappa_1\epsilon_2-\kappa_2\epsilon_1}{
\kappa_1\epsilon_2+\kappa_2\epsilon_1} {\rm e}^{-\kappa_1(x+x')} \Bigg) .
\label{27}
\end{eqnarray}

(ii) If $x<0$ and $x'>0$,
\be \label{28}
D_{xx} = - \frac{4\pi\hbar (cq)^2}{\omega^2} \frac{1}{\kappa_1\epsilon_2
+\kappa_2\epsilon_1} {\rm e}^{\kappa_2 x - \kappa_1 x'} .
\ee

(iii) The case $x>0$ and $x'<0$ is deducible from Eq. (\ref{28}) by using
the symmetry relation (\ref{16}).

(iv) If $x,x'<0$, considering the $1\leftrightarrow 2$ media
exchange symmetry, we obtain from Eq. (\ref{27}) that
\begin{eqnarray} 
D_{xx} & = & \frac{4\pi\hbar c^2}{\omega^2\epsilon_2} \delta(x-x')
- \frac{2\pi\hbar (cq)^2}{\omega^2\epsilon_2\kappa_2} \Bigg(
{\rm e}^{-\kappa_2\vert x-x'\vert} \nonumber \\ & &
\quad + \frac{\kappa_2\epsilon_1-\kappa_1\epsilon_2}{
\kappa_2\epsilon_1+\kappa_1\epsilon_2} {\rm e}^{\kappa_2(x+x')} \Bigg) .
\label{29}
\end{eqnarray}

The symmetrized surface charge correlation function (\ref{2}) is expressible
in terms of the fluctuations of the symmetrized electric field $xx$-components
by using relation (\ref{1}).
These electric-field fluctuations are related to the $xx$ elements of the
Green function tensor via Eq. (\ref{24}).
The terms proportional to $\delta(x-x')$ in Eqs. (\ref{27}) and (\ref{29})
can be ignored since they originate from the short-distance terms
proportional to $\delta({\bf r}-{\bf r}')$ which do not play any role in
the large-distance asymptotic. 
Since the combination
\begin{eqnarray}
D_{xx}(0^+,0^+) + D_{xx}(0^-,0^-) - 2 D_{xx}(0^+,0^-) \nonumber \\
= - \frac{4\pi\hbar (cq)^2}{\omega^2} \frac{1}{\kappa_1\epsilon_2
+\kappa_2\epsilon_1} \left( \frac{\epsilon_2}{\epsilon_1} 
+ \frac{\epsilon_1}{\epsilon_2} - 2 \right) , \label{30}
\end{eqnarray}
we finally arrive at the quantum result
\be \label{31}
\beta S_{\rm qu}(t,q) = \int_{-\infty}^{\infty} 
\frac{{\rm d}\omega}{\omega} {\rm e}^{-{\rm i}\omega t} 
{\rm Im}\, f(\omega) ,
\ee
where, in the retarded regime, $f(\omega)\equiv f_{\rm qu}^{(\rm r)}$ with
\begin{eqnarray} 
f_{\rm qu}^{(\rm r)} & = & \frac{q^2}{4\pi^2} g(\omega) 
\frac{1}{\kappa_1(\omega,q)\epsilon_2(\omega) +
\kappa_2(\omega,q)\epsilon_1(\omega)} \nonumber \\
& & \times \frac{[\epsilon_1(\omega) - \epsilon_2(\omega)]^2}{
\epsilon_1(\omega) \epsilon_2(\omega)} . \label{32}
\end{eqnarray}
In the non-retarded case, $f(\omega)\equiv f_{\rm qu}^{(\rm nr)}$ is
obtained from (\ref{32}) by setting the speed of light $c\to\infty$.
According to Eq. (\ref{26}), the inverse lengths $\kappa_1=\kappa_2=q$
in this limit, so that
\be \label{33}
f_{\rm qu}^{(\rm nr)} = \frac{q}{4\pi^2} g(\omega)
\left[ \frac{1}{\epsilon_1(\omega)} + \frac{1}{\epsilon_2(\omega)}
- \frac{4}{\epsilon_1(\omega)+\epsilon_2(\omega)} \right] .
\ee

\section{4. ANALYSIS OF THE RESULTS}

\subsection{4.1. Jellium conductor in vacuum}
We first considered the previously studied configuration of the jellium
conductor of the dielectric function  $\epsilon_1(\omega)$ given by 
the Drude formula (\ref{9}) in contact with the vacuum of 
$\epsilon_2(\omega)=1$ \cite{SJ08,JS09a}.

In the non-retarded regime, using the Weierstrass theorem
\be \label{34}
\lim_{\eta\to 0^+} \frac{1}{x\pm {\rm i}\eta} = 
{\cal P}\left( \frac{1}{x} \right) \mp {\rm i}\pi\delta(x)
\ee 
(${\cal P}$ denotes the Cauchy principal value) in the representation 
(\ref{33}), the integration over frequency (\ref{31}) leads the previous 
result (\ref{10}).
This result is valid for intermediate distances $R$ on the interface, 
given by the inequalities $\lambda_{\rm ph} \ll R \ll c/\omega_p$,
where $\lambda_{\rm ph}\propto \beta\hbar c$ stands for the thermal
de Broglie wavelength of photon and $c/\omega_p$ is the wavelength
of electromagnetic waves emitted by charge oscillations at frequency
$\omega_p$.

The analysis of the frequency integral (\ref{31}) with the retarded 
function (\ref{32}) is much more complicated \cite{SJ08}. 
At specific values of $\omega$, given by the surface-plasmon dispersion 
relation
\be \label{35}
q^2 (\epsilon_1 + \epsilon_2) - \frac{\omega^2}{c^2} 
\epsilon_1\epsilon_2 = 0,
\ee
the integration passes in the infinitesimal vicinity of a singularity.
The final result for the small $q$-expansion is of the classical static 
type
\be \label{36}
\beta S_{\rm qu}^{(\rm r)}(t,q) = \frac{q}{4\pi} + o(q) .
\ee
In other words, for both static and time-dependent surface charge
correlation functions, the inclusion of retardation effects causes
the quantum prefactor to take its universal static classical form
\be \label{37}
h_{\rm qu}^{(\rm r)}(t) = h_{\rm cl}^{(\rm r)}(0) = 
- \frac{1}{8\pi^2} .
\ee
This formula is valid in the strictly asymptotic region of distances $R$
on the interface, given by the inequality $c/\omega_p\ll R$.
Note that it holds
\be \label{38}
h_{\rm cl}^{(\rm r)}(0) = h_{\rm cl}^{(\rm nr)}(0) \equiv h_{\rm cl}(0) .
\ee
This relation is of general validity due to the Bohr-van Leeuwen theorem 
\cite{Bohr,Leeuwen}.

The same result (\ref{37}) was derived by another (partially microscopic) 
method, based on the analysis of the collective vibration modes 
of the system \cite{JS09a}.

\subsection{4.2. General configuration}
For an arbitrary configuration of the plane contact between two distinct media, 
the small-$q$ analysis of the surface charge correlation function (\ref{31})
was done in Ref. \cite{JS09b} by using general analytic properties of 
dielectric functions in the complex frequency upper half-plane and contour
integration techniques. 

We start with the simpler static case $t=0$. 
The formula (\ref{31}) reads
\be \label{39}
\beta S_{\rm qu}(0,q) = \int_{-\infty}^{\infty} 
\frac{{\rm d}\omega}{\omega} {\rm Im}\, f(\omega) ,
\ee
where the function $f(\omega)$ is given by (\ref{32}) in the retarded regime 
and by (\ref{33}) in the non-retarded regime.
For a real frequency $\omega$, the symmetry relations
$\epsilon^*(\omega)=\epsilon(-\omega)$ and $\kappa^*(\omega)=\kappa(-\omega)$
imply $f^*(\omega)=f(-\omega)$, i.e.
\be \label{40}
{\rm Re}\, f(\omega) = {\rm Re}\, f(-\omega) , \qquad
{\rm Im}\, f(\omega) = - {\rm Im}\, f(-\omega) .
\ee
As $\omega\to 0$, ${\rm Im}\, f(0) = 0$ and 
\be \label{41}
{\rm Re}\, f(0) = \frac{q}{4\pi^2} \left[
\frac{1}{\epsilon_1(0)} + \frac{1}{\epsilon_2(0)}
- \frac{4}{\epsilon_1(0)+\epsilon_2(\omega)} \right] .  
\ee

\begin{figure} [h,t]
\begin{center} 
\includegraphics[width=80mm,clip]{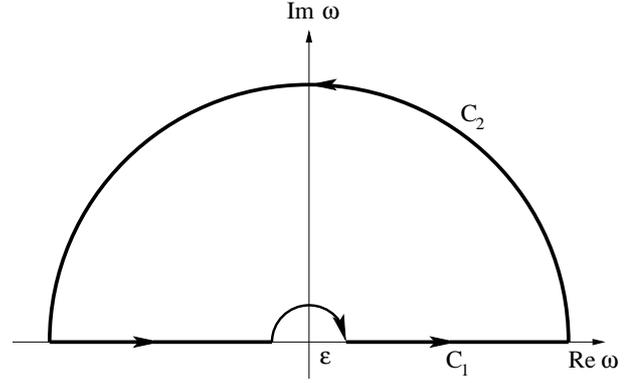}
\end{center}
\vspace{-2mm}
\caption{The contour in the complex frequency plane for $t=0$.}
\end{figure}

Let $C_1$ be the path following the real axis, except it goes around
the origin $\omega=0$ in a small semicircle in complex upper half-plane
whose radius $\epsilon$ tends to zero and $C_2$ be the closure of $C_1$
by a semicircle at infinity (see Fig. 2).
The integration along the real axis in Eq. (\ref{39}) is expressible 
in terms of the path integration over the closed contour $C=C_1\cup C_2$ 
as follows
\be \label{42}
\int_{-\infty}^{\infty} \frac{{\rm d}\omega}{\omega} {\rm Im}\, f(\omega)
= \pi f(0) + {\rm Im} \oint_C \frac{{\rm d}\omega}{\omega} f(\omega) .
\ee
The integral over the contour $C$ can be evaluated by using the residue 
theorem at poles $\{ \omega_j \}$ of the function $f(\omega)$ in the
$\omega$ upper half-plane bounded by $C$,
\be \label{43}
{\rm Im} \oint_C \frac{{\rm d}\omega}{\omega} f(\omega)
= 2\pi \sum_j \frac{{\rm Res}(f,\omega_j)}{\omega_j} ,
\ee 
provided that ${\rm Res}(f,\omega_j)/\omega_j$ is real 
(which will be the case); Res denotes the residue.
Both the retarded (\ref{32}) and non-retarded (\ref{33}) versions
of the $f$-function contain $g(\omega)$ defined in (\ref{11}).
Since $g(\omega)$ can be expanded in $\omega$ as \cite{Gradshteyn}
\be \label{44}  
g(\omega) = 1 + \sum_{j=1}^{\infty} \frac{2\omega^2}{\omega^2+\xi_j^2} ,
\qquad \xi_j = \frac{2\pi}{\beta\hbar} j ,
\ee
it has in the upper half-plane an infinite sequence of simple poles 
at the imaginary Matsubara frequencies
\begin{equation} \label{45}
\omega_j = i \xi_j , \qquad {\rm Res}(g,\omega_j) = \omega_j 
\quad (j=1,2,\ldots) .
\end{equation}
Using general analytic properties of dielectric functions it was shown 
in Ref. \cite{JS09b} that, in both retarded and non-retarded regimes,
these are the only poles of the $f$-function inside the contour $C$.
The static correlation function (\ref{39}) is therefore expressible as follows
\begin{eqnarray} 
\beta S_{\rm qu}(0,q) & = & \frac{q}{4\pi} 
\left[ \frac{1}{\epsilon_1(0)} + \frac{1}{\epsilon_2(0)} 
- \frac{4}{\epsilon_1(0)+\epsilon_2(0)} \right] \nonumber \\ & &
+ F(0,q) , \label{46} \\
F(0,q) & = & 2\pi \sum_{j=1}^{\infty} \frac{{\rm Res}(f,i\xi_j)}{i\xi_j} . 
\label{47}
\end{eqnarray}
The first term on the rhs of Eq. (\ref{46}), linear in $q$, is 
independent of $\beta\hbar$ and $c$, the $q$-dependence of the
static function $F(0,q)$ depends on the considered 
(retarded or non-retarded) regime.

In the retarded case (\ref{32}), we have
\begin{eqnarray}
F_{\rm qu}^{(\rm r)}(0,q) & = &  \frac{q^2}{2\pi} \sum_{j=1}^{\infty}
\frac{1}{\kappa_1(i\xi_j) \epsilon_2(i\xi_j) +
\kappa_2(i\xi_j) \epsilon_1(i\xi_j)} \nonumber \\ & & \times
\frac{[\epsilon_1(i\xi_j)-\epsilon_2(i\xi_j)]^2}{\epsilon_1(i\xi_j)
\epsilon_2(i\xi_j)} . \label{48}
\end{eqnarray}
We are interested in the limit $q\to 0$ for which
$\kappa_{1,2}(i\xi_j) \sim \xi_j \epsilon_{1,2}^{1/2}(i\xi_j)$
($\epsilon_{1,2}(i\xi_j)$ are real).
Since $\xi_j\propto j$ and $\epsilon(i\xi_j)-1 = O(1/j^2)$ in the limit
$j\to\infty$, the sum in (\ref{48}) converges.
This means that the function $F_{\rm qu}^{(\rm r)}(0,q)$, being of 
the order $O(q^2)$, becomes negligible in comparison with the first 
term in Eq. (\ref{46}) when $q\to 0$.
We find the static $h$-prefactor associated with the asymptotic decay to be
\be \label{49}
h_{\rm qu}^{(\rm r)}(0) = - \frac{1}{8\pi^2} 
\left[ \frac{1}{\epsilon_1(0)} + \frac{1}{\epsilon_2(0)} 
- \frac{4}{\epsilon_1(0)+\epsilon_2(0)} \right] .
\ee
Since this expression does not depend on the temperature and $\hbar$,
its classical $\beta\hbar\to 0$ limit is the same, i.e.
$h_{\rm cl}^{(\rm r)}(0) = h_{\rm cl}^{(\rm nr)}(0) \equiv h_{\rm cl}(0)$ with 
\be \label{50}
h_{\rm cl}(0) = - \frac{1}{8\pi^2} 
\left[ \frac{1}{\epsilon_1(0)} + \frac{1}{\epsilon_2(0)} 
- \frac{4}{\epsilon_1(0)+\epsilon_2(0)} \right] .
\ee
This classical result was derived independently in the non-retarded 
regime by using the linear response theory combined with the electrostatic 
method of images \cite{JS09b}. 
We see that from the dielectric functions of the two media only their values
at zero frequency appear; we recall that $\epsilon(0)\to i\infty$ 
for conductors, $\epsilon(0) = \epsilon_0>1$ for dielectrics and 
$\epsilon(0) = 1$ for vacuum.
The $h$-prefactor is nonzero for an arbitrary configuration of
different media, except for the special case of two conductors.

In the non-retarded case (\ref{33}), we have
\begin{eqnarray}
F_{\rm qu}^{(\rm nr)}(0,q) & = & \frac{q}{2\pi} \sum_{j=1}^{\infty}
\left[ \frac{1}{\epsilon_1(i\xi_j)} + \frac{1}{\epsilon_2(i\xi_j)}
\right. \nonumber \\ & & \left. 
- \frac{4}{\epsilon_1(i\xi_j)+\epsilon_2(i\xi_j)} \right] . \label{51}
\end{eqnarray}
It is evident from the asymptotic behavior $\epsilon(i\xi_j)-1 = O(1/j^2)$ 
that the sum in (\ref{51}) converges.
The function $F_{\rm qu}^{(\rm nr)}(0,q)$ is thus of the order $O(q)$ and
$h_{\rm qu}^{(\rm nr)}(0)$ is a complicated function of the temperature.

A more laborious analysis is needed for the time-dependent case $t\ne 0$
\cite{JS09b}.
The final result is
\be \label{52}
h_{\rm qu}^{(\rm r)}(t) = - \frac{1}{8\pi^2} 
\left[ \frac{1}{\epsilon_1(0)} + \frac{1}{\epsilon_2(0)} 
- \frac{4}{\epsilon_1(0)+\epsilon_2(0)} \right] .
\ee
As before, the quantum time-dependent prefactor to the asymptotic decay
takes, for any temperature, its classical static form (\ref{50}).

\section{5. CONCLUSION}
Although the present study is rather technical, its main result (\ref{52})
is of importance.
We do not understand which are {\em physical} reasons that the inclusion
of retardation effects causes the asymptotic decay of time-dependent
quantum surface charge correlation functions to take its static classical
form, independent of the temperature factor $\beta\hbar$ and the speed of
light $c$.

This is one of the rare phenomena in the Condensed Matter physics when
retardation (relativistic) effects play an essential role. 
\\

\noindent ACKNOWLEDGMENT: L. \v{S}. is grateful to LPT for very
kind invitations and hospitality.
The support received from the MISGAM program of the European Science 
Foundation, Grant VEGA No. 2/0113/2009 and CE-SAS QUTE is acknowledged.
\\

\end{document}